\documentclass[twocolumn,showpacs,preprintnumbers,amsmath,amssymb,prl,superscriptaddress]{revtex4-1}
\pdfoutput=1
\usepackage{graphicx}
\usepackage{dcolumn}
\usepackage{color}

\usepackage[low-sup]{subdepth}
\usepackage{multirow}
\usepackage{amsfonts}
\usepackage[english]{babel}
\usepackage[T1]{fontenc}
\usepackage{times}
\usepackage[colorlinks,bookmarks=true,citecolor=blue,linkcolor=red,urlcolor=blue]{hyperref}
\usepackage[tight, FIGTOPCAP, hang, raggedright, nooneline]{subfigure}

\usepackage{verbatim}
\usepackage{bm}
\usepackage{mathrsfs}

\newcommand{\ket}[1]{\left|#1\right>}

\newcommand{\para}[1]{\left(#1\right)}
\newcommand{\abs}[1]{\left|#1\right|}

\newcommand{\COMMENT}[1]{}

\begin{document}
\title{Non-Abelian phases in two-component $\nu=2/3$ fractional quantum Hall states: Emergence of Fibonacci anyons}
\author{Zhao Liu}
\affiliation{Department of Electrical Engineering, Princeton University, Princeton, New Jersey 08544,
USA}
\author{Abolhassan Vaezi}
\email{vaezi@cornell.edu}
\author{Kyungmin Lee}
\author{Eun-Ah Kim}
\affiliation{Department of Physics, Cornell University, Ithaca, New York 14853, USA}

\begin{abstract} Recent theoretical insights into the possibility of non-Abelian phases in $\nu=2/3$  fractional quantum Hall states revived the interest in the numerical phase diagram of the problem. We investigate the effect of various kinds of two-body interlayer couplings on the $(330)$ bilayer state and exactly solve the Hamiltonian for up to $14$ electrons on sphere and torus geometries. We consider interlayer tunneling, short-ranged repulsive/attractive pseudopotential interactions and Coulomb repulsion. We find a 6-fold ground-state degeneracy on the torus when the interlayer hollow-core interaction is dominant. To identify the topological nature of this phase we measure the orbital-cut entanglement spectrum, quasihole counting, topological entanglement entropy, and wave-function overlap. Comparing the numerical results to the theoretical predictions, we interpret this 6-fold ground-state degeneracy phase to be the non-Abelian bilayer Fibonacci state.
\end{abstract}

\maketitle

{\em Introduction.} There is a surging interest in the search for non-Abelian (NA) anyons in topological states of matter~\cite{Nayak2008,alicea2012review,vaezi2014b,Lindner2012,Clarke2013,Cheng2012,Vaezi2013,Barkeshli2013defect}. NA anyons register protected quantum memory and the stored information can be processed through braiding the world line of NA anyons. This comprises the basic notion of building fault-tolerant quantum computers. Two famous examples are Ising and Fibonacci anyons which have been conjectured to emerge in $\nu=5/2$ and $\nu=12/5$ fractional quantum Hall (FQH) states, respectively \cite{moore1991,wen1991prl,read1999}. In particular Fibonacci anyons are of interest for universal quantum computation. However, there is no experimental observation of these exotic quasiparticles in FQH systems to date. Another approach for realizing NA anyons is through coupling Abelian states via various types of interactions to drive phase transition to NA topological phases~\cite{ardonne1999, Cappelli2001, wen2000, fradkin1999, papic2010, peterson2010}. In a realistic experimental situation, some of the interactions considered in theoretical investigations are indeed relevant~\cite{Eisenstein1990,suen1994,manoharan1996,lay1997,Shabani2010,Du2009,Bolotin2009,Dean2011}. Pursuing this direction further can potentially lead to a novel venue for realizing NA anyons.

In this Rapid Communication, we focus on the two-component FQH system at $\nu=2/3$, whose parent state consists of two decoupled $\nu=1/3$ Laughlin states. The two components may describe physical spins~\cite{Eisenstein1990}, spatial (physical layer) degrees of freedom, e.g., in the wide quantum well realization of the $2/3$ state~\cite{suen1994,manoharan1996,lay1997,Shabani2010}, or valley indices in graphene-like systems~\cite{Du2009,Bolotin2009,Dean2011}. For convenience, we refer to all these various realizations of two-component systems as the ``{\em $2/3$ bilayer}'' state.
Early studies of the $2/3$ bilayer state focused on various possible Abelian phases~\cite{Halperin1983} and the experimental observation of phase transitions~\cite{Eisenstein1990,suen1994}. In particular McDonald and Haldane numerically established the phase diagram of the system in the presence of interlayer tunneling and the bare Coulomb interaction projected to the lowest Landau level (LLL)~\cite{McDonald1996} and confirmed the possibility of various Abelian phases.
Recent analytical insights into the possibility of NA phases in the $2/3$ bilayer state~\cite{vaezi2014b,rezayi2010,barkeshli2010prl,Vaezi2014a,Mong2014,Ardonne2002,Bonderson2008} brought this seemingly closed problem into new focus. However, the results of Ref.~\cite{McDonald1996} imply that microscopic realizations of these NA phases require perturbation to the model Hamiltonian consisting of interlayer tunneling and the bare Coulomb interaction projected to the LLL. In fact such perturbations may already exist in experimental systems.

Therefore, the recent proposals~\cite{vaezi2014b,rezayi2010,barkeshli2010prl,Vaezi2014a,Mong2014,Ardonne2002,Bonderson2008}
beg for revisiting the $2/3$ bilayer problem numerically considering the modified interlayer Coulomb interaction and interlayer tunneling.
Any two-body interaction projected to the LLL can be expanded in terms of the so-called Haldane pseudopotentials, $\mathcal{V}_m$~\cite{Haldane1983,Davenport2012}, which are projector operators of two electrons to the state with relative angular momentum $m$. Modifying the Coulomb interaction amounts to changing the coefficient of the pseudo-potentials. Here, we modify the interlayer Coulomb interaction by changing its dominant components, namely $\mathcal{V}^{\textrm{inter}}_0$ and $\mathcal{V}^{\textrm{inter}}_1$ (also known as hollow-core interaction).  We solve the resulting Hamiltonian for up to $14$ electrons using the exact-diagonalization (ED) method and utilize a variety of numerical measurements to establish the nature of different topological phases achieved through varying coupling parameters.  We obtain a wide range of parameters where enhancing $\mathcal{V}^{\textrm{inter}}_1$ and suppressing $\mathcal{V}^{\textrm{inter}}_0$ components of the Coulomb interaction drives phase transition to a NA state with Fibonacci anyons.

\noindent {\em Model.} We start with two $\nu=1/3$ Laughlin states that are coupled through a number of distinct interlayer interactions. Let us now consider a torus with $L_x,L_y$ dimensions. In the Landau gauge ${\bf A}=B(0,x)$, single-particle states in the LLL are labeled by their momentum along the $y$ direction quantized as $k_y=2\pi n/L_y$. 
The parent state, the $(330)$ Halperin state \cite{Halperin1984}, is the exact eigenstate of the following $\mathcal{V}_1$ Haldane's pseudopotential
\begin{eqnarray}
\mathcal{V}_1^{\textrm{intra}}=\sum_{\sigma=\uparrow,\downarrow}\sum_{n,r,s}V^{\para{1}}_{r,s} c^\dag_{n+r,\sigma}  c^\dag_{n+s,\sigma} c_{n,\sigma} c_{n+r+s,\sigma},
\end{eqnarray}
where $\sigma=\uparrow,\downarrow$ denotes the layer index, $c_{n,\sigma}$ annihilates an electron with $k_y=2\pi n/L_y$ momentum, $V^{\para{1}}_{r,s}=\frac{\kappa^3}{\sqrt{2\pi}} \para{r^2-s^2}e^{-\kappa^2\frac{r^2+s^2}{2}}$, and $\kappa=2\pi/L_y$.

We investigate the effects of interlayer coupling by considering both tunneling and interlayer two-body interaction of Haldane's $\mathcal{V}_0$ and $\mathcal{V}_1$ pseudo-potentials. The tunneling Hamiltonian is
\begin{eqnarray}
\mathcal{H}_{t}=-t_{\perp}\sum_{n} c_{n}^\dag \sigma_x c_{n},
\end{eqnarray}
where $c_{n}^{\rm T}\equiv\para{c_{n,\uparrow},c_{n,\downarrow}}$. The two-body interactions are
\begin{eqnarray}
\mathcal{V}_0^{\textrm{inter}}=U_0 \sum_{\sigma;n,r,s} V^{\para{0}}_{r,s} c^\dag_{n+r,\sigma} c^\dag_{n+s,-\sigma} c_{n,-\sigma}c_{n+r+s,\sigma},
\end{eqnarray}
where $V^{\para{0}}_{r,s}=\frac{\kappa}{\sqrt{2\pi}} e^{-\kappa^2\frac{r^2+s^2}{2}}$ and
\begin{eqnarray}
\mathcal{V}_1^{\textrm{inter}}=U_1 \sum_{\sigma; n,r,s} V^{\para{1}}_{r,s} c^\dag_{n+r,\sigma}  c^\dag_{n+s,-\sigma} c_{n,-\sigma} c_{n+r+s,\sigma}.
\end{eqnarray}
In order to address the stability of the phases that we obtain with this model we also consider the effect of interlayer Coulomb interaction in the limit of negligible interlayer separation perturbed by the above terms.

{\em Candidate phases.} Below, we consider two categories of possible states at $\nu=2/3$: Abelian and non-Abelian, each with three members. Table \ref{Tab1} summarizes some of their defining properties. (i) The $(330)$ state, which is simply two decoupled $\nu=1/3$ Laughlin states.
This phase has $9$-fold degenerate ground states on the torus. (ii) $\overline{1/3}$: The particle-hole(PH) conjugate of the $\nu=1/3$ Laughlin state~\cite{Jain1989,Jainbook}. (iii) Layer singlet state: The nonchiral $(112)$ Halperin state~\cite{Halperin1984,Jain1989,wenbook,Jainbook}. 
(iv) The bilayer Fibonacci phase~\cite{vaezi2014b}, which is a non-Abelian state described by $SU(3)_2$ Chern-Simons field theory and contains $6$ distinct quasiparticles, three of which are non-Abelian with $d=\frac{1+\sqrt{5}}{2}$ quantum dimension.
(v) The interlayer (I.~L.) Pfaffian state~\cite{Ardonne2002}, which has $9$ distinct quasiparticles and its wave-function can be written as the product of the $(221)$ Halperin state and a Pfaffian wave-function \cite{Ardonne2002}.
(vi) The fermionic $(M=1)$ $Z_4$ Read-Rezayi (RR) state~\cite{rezayi2010,barkeshli2010prl,Lee2015}. This non-Abelian state with $15$ distinct quasiparticles is described by $Z_4 \times U(1)_6$ edge conformal field theory. A unique feature of this state is the emergence of a non-Abelian quasihole excitation with $-e/6$ fractional charge.

\begin{table}
  \centering
  \begin{tabular}{@{} cccccc @{}}
  Phase &$c$& {\rm GSD} & $\mathcal{S}$&$\mathcal{D}$ & OES~~~~   \\
    \hline
    $(330)$ &$2$&$9$ & $3$&$ 3$ & $~~~1,2,5,10,20,\cdots$  \\
    $\overline{1/3}$ &$ 0$&$3$ &$0$& $ \sqrt{3}$ & $1,1,2,3,5,\cdots$  \\
    $(112)$ & $0$&$3$ & $1$&$ \sqrt{3}$ &  \\
    Fibonacci &$14/5$& $6$ &$3$& $\sqrt{\frac{3}{2}\para{5+\sqrt{5}}}$ & ~$1,1,3,6,13,\cdots$   \\
    I.~L. Pfaffian & $5/2$&$9$ &$3$& $\sqrt{12}$ & ~~$1,2,6,13,\cdots$   \\
    $Z_4$ RR &$2$& $15$ &$3$& $6$ & ~$1,1,3,6,13,\cdots$  \\
  \end{tabular}
  \caption{Total central charge $c$, ground-state degeneracy (GSD), the sphere shift $\mathcal{S}=\nu^{-1}N_e-N_{\Phi}$, total quantum dimension $\mathcal{D}$, and orbital-cut entanglement spectrum (OES) counting for $l=0,1,2,3,4,\cdots$ ($l$ is the angular momenta relative to the ground state) for various topological phases at $\nu=2/3$. For NA states, the OES counting generally depends on the number of electrons $N_e^A$ and the pseudospin $S_z^A$ in region $A$. Here we only show the case when $N_e^A/2$ ($N_e^A/4$) is an integer and $S_z^A=0$ for the Fibonacci and I.~L. Pfaffian states ($Z_4$ RR state). Although the countings of the bilayer Fibonacci state with $N_e^A=0$ (mod 2) and $Z_4$ RR state with $N_e^A=0$ (mod 4) are equal for $l\leq 4$, they are different for other values of $N_e^A$. The counting of the $(112)$ state is the convolution of the counting of two counter-propagating $U(1)$ modes.
}
  \label{Tab1}
\end{table}

\begin{figure}
\centerline{\includegraphics[width=0.85\linewidth] {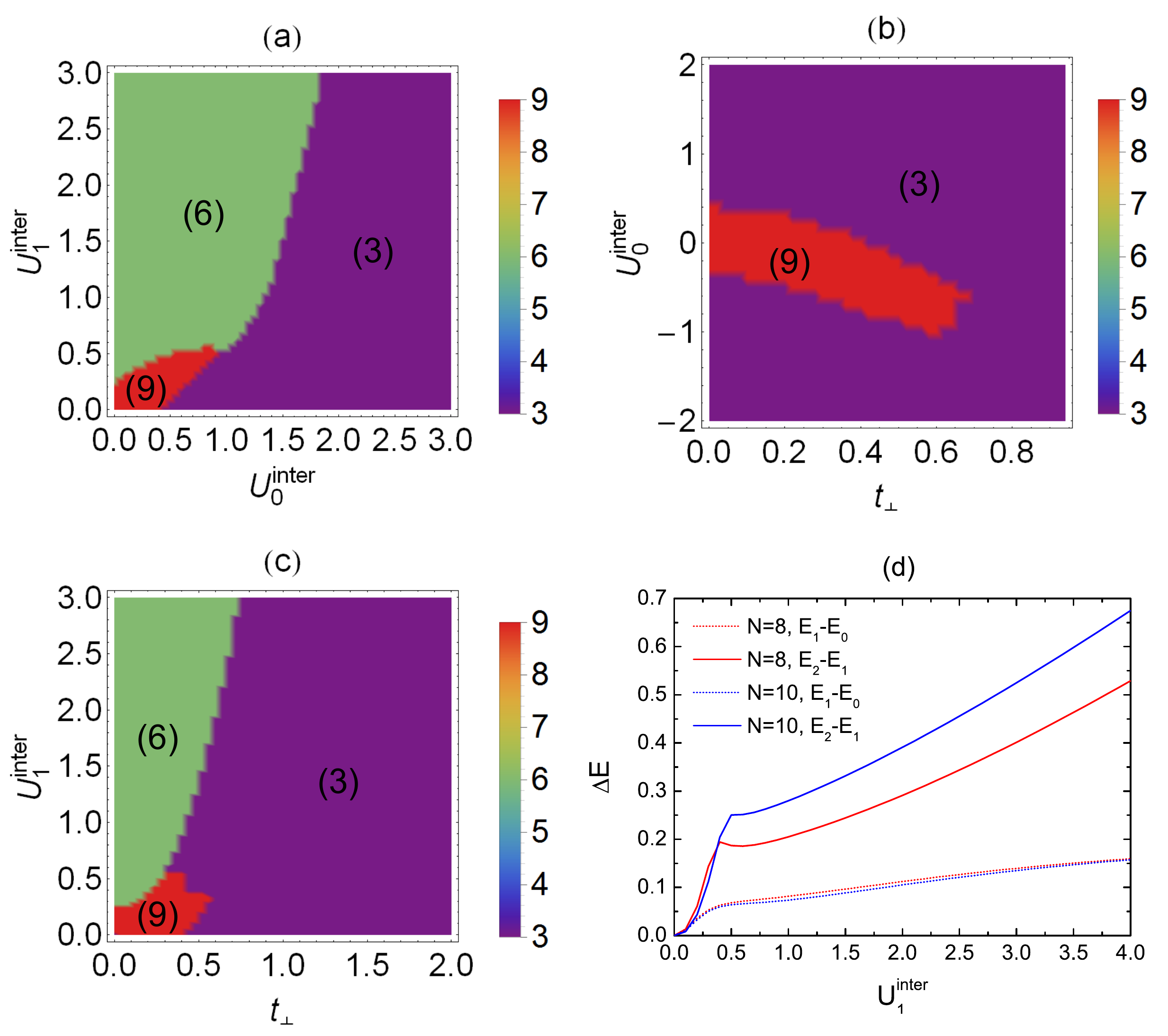}}
\caption{\label{fg:phasediagram} (Color online) The phase diagram based on GSD of the Hamiltonian $H=\mathcal{V}_1^{\textrm{intra}}+U_0^{\textrm{inter}}\mathcal{V}_0^{\textrm{inter}}+U_1^{\textrm{inter}}\mathcal{V}_1^{\textrm{inter}}+\mathcal{H}_{t}$. The numbers in the parentheses indicate GSD. (a)-(c) The GSD for $N_e=8$ particles for (a) $t_{\perp}=0$, (b) $U_1^{\textrm{inter}}=0$, and (c) $U_0^{\textrm{inter}}=0$. (d) The GSD as a function of $U_1^{\textrm{inter}}$ for $N_e=8$ and $10$ electrons with $U_0^{\textrm{inter}}=t_{\perp}=0$. $E_0,E_1,E_2$ are the lowest three energy levels in the $K<N_\Phi/3$ sectors. One can see that with the increase of $U_1^{\textrm{inter}}$, $E_2-E_1\gg E_1-E_0$, implying a total $6$-fold degeneracy.
}
\end{figure}

{\em Numerical results.}
Now we numerically search for possible phases by exact diagonalization of our model Hamiltonian. To mitigate the limitations of finite-size system study, we make measurements of various characteristics of topological phases using both toroidal and spherical geometries.
The number of electrons $N_e$ and total number of quantum fluxes $N_{\Phi}$ are related through the $N_{\Phi}=\frac{3}{2}N_e$ relation on the torus and on the sphere $N_{\Phi}=\frac{3}{2}N_e-\mathcal{S}$, where $\mathcal{S}$ is the shift of a FQH state on the sphere~\cite{Wen1992prl}.
Good quantum numbers that can be used to label energy states are the total center-of-mass momentum $K$ on the torus and angular momentum $L_z$ on the sphere. In the absence of the interlayer tunneling, the total pseudospin $S_z=\frac{1}{2}(N_{\uparrow}-N_{\downarrow})$ is also a good quantum number, where $N_{\uparrow,(\downarrow)}$ is the electron number in the upper (lower) layer.

Below we present the resulting phase diagram and the measurements characterizing the phases.

(a) {\em Ground-state degeneracy.} The GSD on the torus is significant as it equals the number of distinct anyon excitations~\cite{wen1992}. Figure~\ref{fg:phasediagram} shows the phase diagram of the interaction $H=\mathcal{V}_1^{\textrm{intra}}+U_0^{\textrm{inter}}\mathcal{V}_0^{\textrm{inter}}+U_1^{\textrm{inter}}\mathcal{V}_1^{\textrm{inter}}+\mathcal{H}_{t}$ exhibiting several phases with $3$-fold (purple), $6$-fold (green), and $9$-fold (red) GSD phases. Of these the $9$-fold GSD region in Figs.~\ref{fg:phasediagram}(a) and \ref{fg:phasediagram}(b) is smoothly connected to the parent state, so it is most likely the $(330)$ phase. For the $3$-fold GSD region, there are two Abelian candidates (see Table \ref{Tab1}) which can be differentiated by their different shifts on sphere, $\mathcal{S}$. The most tantalizing is the $6$-fold GSD region, for which the NA bilayer Fibonacci state is the only candidate to the best of our knowledge. Figures~\ref{fg:phasediagram}(a)-\ref{fg:phasediagram}(c) show that dominant $\mathcal{V}^{\textrm{inter}}_1$ is required for the $6$-fold GSD phase. Nevertheless, this phase is stable against subdominant tunneling and $\mathcal{V}_0^{\textrm{inter}}$ terms. In the rest of this Rapid Communication, we will focus on the $t_{\perp}=0$ limit where we can take advantage of $S_z$ as a good quantum number. Figure~\ref{fg:phasediagram}(d) shows that this degeneracy is stable upon increasing the system size.

\begin{figure}
\centerline{\includegraphics[width=0.85\linewidth] {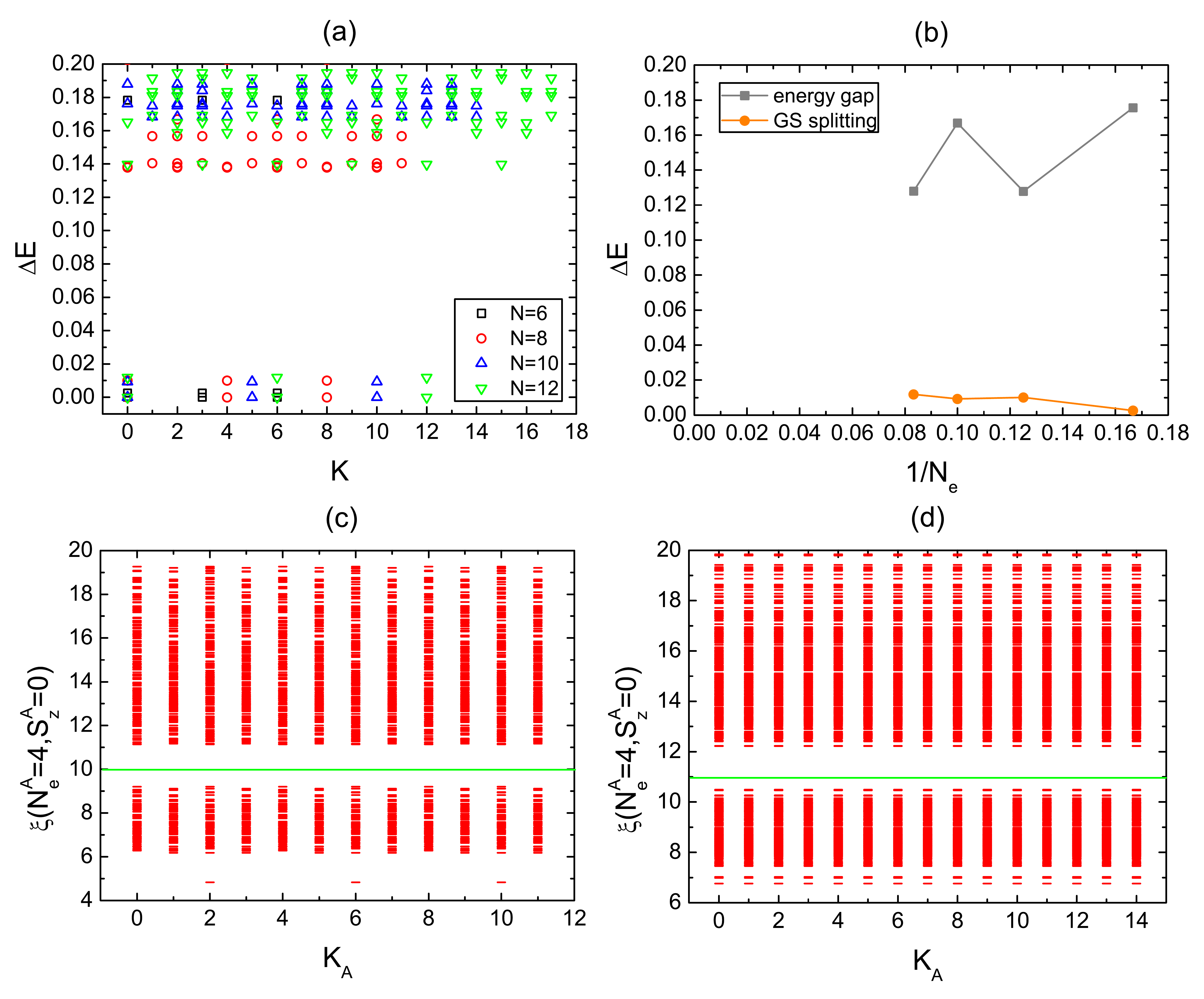}}
\caption{(Color online) 
Here we consider the interaction $H_{\textrm{coulomb}}^{\textrm{intra}}+H_{\textrm{coulomb}}^{\textrm{inter}}+U_0^{\textrm{inter}}\mathcal{V}_0^{\textrm{inter}}+U_1^{\textrm{inter}}\mathcal{V}_1^{\textrm{inter}}$ with $U_0^{\textrm{inter}}=-0.4$ and $U_1^{\textrm{inter}}=0.6$ on the square torus \cite{PS}.
(a) The energy spectrum of $N_e=6,8,10,$ and $12$ electrons in the $S_z=0$ sector. (b) The finite-size scaling of the energy gap and the ground-state splitting in the $S_z=0$ sector. (c) The PES of $N_e=8$ electrons; the counting below the gap (green line) is 1332. (d) The PES of $N_e=10$ electrons; the counting below the gap (green line) is 4875.}
\label{fg:torus}
\end{figure}

\begin{figure}
\centerline{\includegraphics[width=0.85\linewidth] {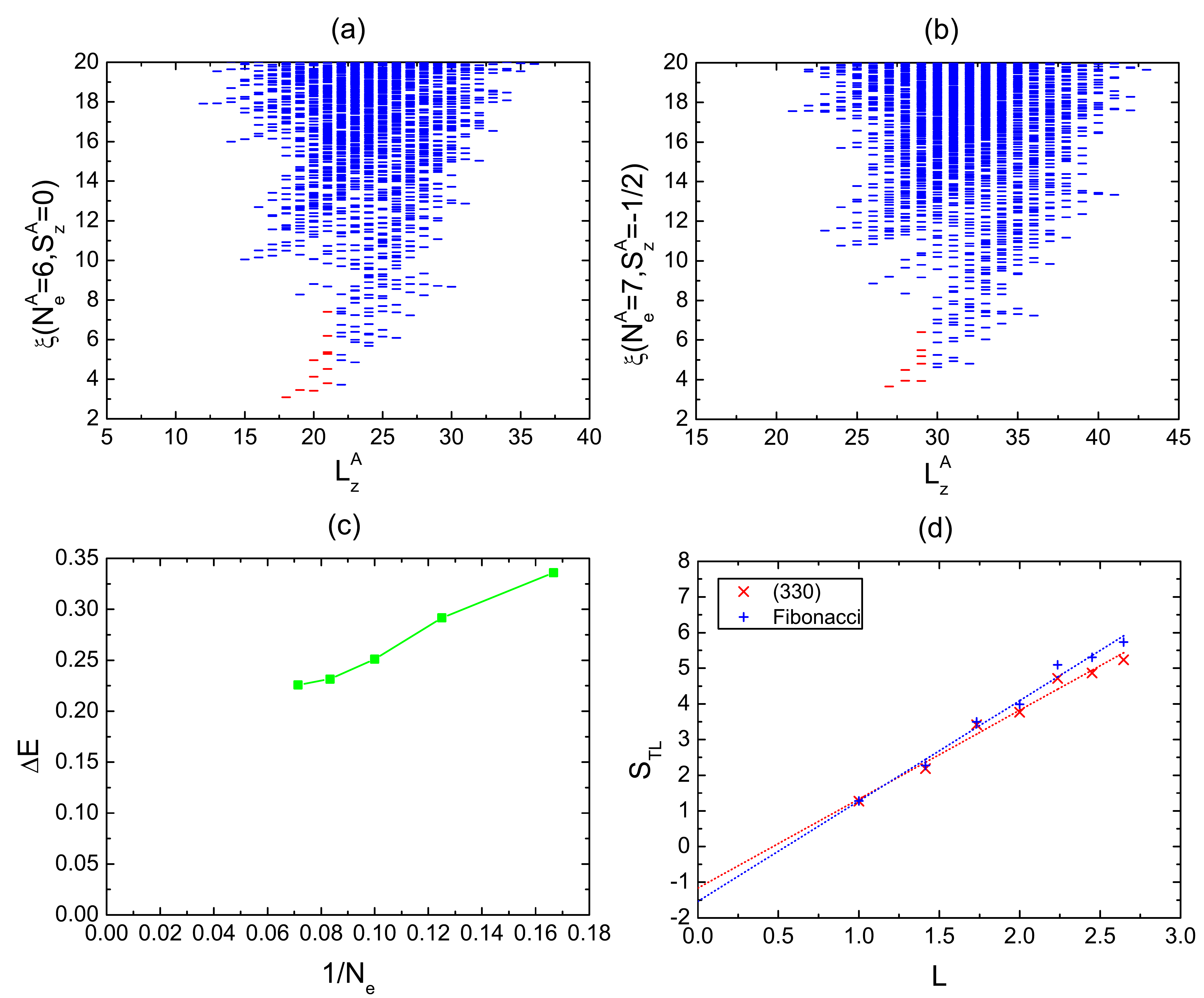}}
\caption{(Color online)
For (a)-(c) we consider the modified Coulomb interaction with $H_{\textrm{coulomb}}^{\textrm{intra}}+H_{\textrm{coulomb}}^{\textrm{inter}}+U_0^{\textrm{inter}}\mathcal{V}_0^{\textrm{inter}}+U_1^{\textrm{inter}}\mathcal{V}_1^{\textrm{inter}}$ with $U_0^{\textrm{inter}}=-0.4$ and $U_1^{\textrm{inter}}=0.6$ on the sphere. (a) and (b) show the OES for $N_e=14$ electrons in the $N_e^A=6,S_z^A=0$ and $N_e^A=7,S_z^A=-1/2$ sectors. The levels consistent with the exclusion rule proposed in the text are colored by red. (c) The finite-size scaling of the energy gap in the $S_z=0$ sector.
(d) The extrapolation of the measured entanglement entropy $S_{A}(L)$ to the thermodynamic limit, $S_{\textrm{TL}}$, for different boundary lengths $L$, using the method in Ref.~\cite{tee1} for the unperturbed $(330)$ state and the 6-fold GSD phase in the large-$\mathcal{V}_1^{\textrm{inter}}$ limit.}
\label{fg:sphere}
\end{figure}

To further address the stability of the $6$-fold GSD phase against higher order pseudopotentials which must exist in any realistic setting, we study the $H=H_{\textrm{coulomb}}^{\textrm{intra}}+H_{\textrm{coulomb}}^{\textrm{inter}}+U_0^{\textrm{inter}}\mathcal{V}_0^{\textrm{inter}}+U_1^{\textrm{inter}}\mathcal{V}_1^{\textrm{inter}}$ Hamiltonian with $U_0^{\textrm{inter}}<0$. We consider this particular form, because the $\mathcal{V}_0^{\textrm{inter}}$ component of the bare Coulomb interaction is larger than its $\mathcal{V}_1^{\textrm{inter}}$ component.
Intriguingly, Fig.~\ref{fg:torus}(a) shows that this Hamiltonian can stabilize the $6$-fold GSD phase with a robust degeneracy and a large gap. Moreover, the finite-size scaling of the energy gap and the ground-state splitting [Fig.~\ref{fg:torus}(b)] shows that this phase is probably gapped in the thermodynamic limit. The finite-size scaling of the energy gap on the sphere also supports this conclusion [Fig.~\ref{fg:sphere}(c)].

(b) {\em Edge state counting and entanglement spectrum.} The orbital-cut entanglement spectrum (OES)~\cite{Li2008} encodes the degeneracy of the edge excitations with angular momentum $l$ relative to the ground state~\cite{wenbook}. We compute the OES on the sphere which leads to a single boundary.
For a bilayer FQH system without interlayer tunneling, each OES level on the sphere can be labeled by the total electron number $N_e^A$, the total angular momentum $L_z^A$, and the total pseudospin $S_z^A$ in the subregion $A$.

Table \ref{Tab1} lists the expected counting for the candidate phases that can be obtained using the thin torus patterns and generalized exclusion rules~\cite{Bernevig2008,SM}. Care has to be given to the fact that OES counting generally depends on $N_e^A$. For example, in the case of Abelian phases, the OES is independent of $N_e^A$, while for the bilayer Fibonacci state it depends on the parity of $N_e^A$. In the case of the $Z_4$ RR state, the counting depends on $N_e^A$ modulo $4$. In general, it can be shown that the OES counting of a $k-$electron clustered topological phase described by $SU(n)_k$ Chern-Simons theory depends on $N_e^A$ modulo $k$. For the bilayer Fibonacci state, the counting for $N_e^A=2M+1$ is $1,2,5,10,\cdots$. For the $Z_4$ RR state, the counting is $1,2,5,10,\cdots$ ($1,2,6,11,\cdots $) for $N_e^A=4M-1$ ($N_e^A=4M-2$) for large values of $N_e^A$. For finite systems, the counting is slightly different. For instance, when $N_e=6$ the expected counting for the bilayer Fibonacci ($Z_4$ RR) state is $1,1,3,6,12,\cdots$ ($1,2,5,7,\cdots$). For $N_e^A=7$, the expected counting of the $Z_4$ RR state becomes $1,2,4,7,\cdots$.

For the region with 9-fold GSD we obtained OES counting consistent with the Abelian $(330)$ state: $1,2,5,10,20,\cdots$ independently of $N_e^A$. On the other hand, the measured OES counting for the $6$-fold GSD phase is different for the odd and even values of $N_e^A$, and distinct from any Abelian state. The OES counting of the modified interlayer Coulomb interaction with 6-fold GSD is presented in Figs.~\ref{fg:sphere}(a) and \ref{fg:sphere}(b). The OES counting we obtain for the bilayer Fibonacci state dictates the following generalized Pauli exclusion rules: (a) In every three consecutive orbitals, there are at most two electrons.
(b) Each configuration of two electrons with distance more than two orbitals such as (1001) or (10001) is doubly degenerate after assigning spin indices and must be counted twice.  It was shown that these rules give rise to 6-fold GSD on the torus, and result in three NA anyon excitations with $d=\frac{1+\sqrt{5}}{2}$ quantum dimension, hence the Fibonacci anyons~\cite{vaezi2014b}.

The fact that the counting for $l=1$ is unambiguously $1$ for $N_e^A=6$ rules out the I.~L. Pfaffian and $Z_4$ RR states (for both of these states at $l=1$, we expect the counting to be $2$).
Moreover, the fact that the measured OES counting depends on the parity of $N_e^A$ indicates that the $6$-fold GSD phase is a paired state with $k=2$. This is consistent with the $SU(3)_2$ Chern-Simons description of the bilayer Fibonacci state.

(c) {\em quasihole counting.} The degeneracy of the ground-state in the presence of quasihole excitations, usually referred to as quasihole counting, is a telling property of a topological phase~\cite{Read1996,Simon2007}. One way to create quasiholes is by adding additional fluxes to the system. For instance, in the bilayer Fibonacci phase, each additional flux would create two quasiholes. Another way is to use the particle-cut entanglement spectrum (PES) which is achieved by tracing out some electrons in the density matrix associated with the ground-state wave function~\cite{Sterdyniak2011}. Each PES level is labeled by the total electron number $N_e^A$, momentum $K_A$, and pseudospin $S_z^A$ in the subsystem $A$. In Figs.~\ref{fg:torus}(c) and \ref{fg:torus}(d), we observe a clear gap in the PES. The counting below gap is different from the expected counting for the Abelian as well as I.~L. Paffian and $Z_4$ RR states~\cite{SM}, while it is consistent with the generalized Pauli exclusion rules of the bilayer Fibonacci state described in the previous section.

(d) {\em Topological entanglement entropy.} The entanglement entropy of a two-dimensional gapped state follows the $S_{A}(L)=\alpha L -\gamma$ relation, where $L$ is the perimeter of subregion $A$. The subleading constant $\gamma$ is related to the total quantum dimension of the topological phase $\mathcal{D}$, through the $\gamma=\log \mathcal{D}$~\cite{Kitaev2006b,Levin2006} relation. The precise numerical evaluation of the topological entanglement entropy (TEE) $\gamma$ for small systems is usually challenging~\cite{tee1,Zozulya2007,andreastorusTEE,Liu2012}. However, qualitative aspects of our results in Fig.~\ref{fg:sphere}(d) suggest that the 6-fold GSD phase is a NA state. The numerically estimated $\gamma$ for the 6-fold GSD state is larger than the estimated $\gamma$ for the unperturbed $(330)$ state and far from the values of $\gamma$ for the remaining Abelian phases whose $\gamma$'s are expected to be half the value of the $(330)$ state~(see Table \ref{Tab1}). Although this argument suggests the resulting state is NA, it does not rule out the possibility of other NA states.

(e) {\em Wave-function overlap.} We have also computed the squared overlap of the ground state of the 6-fold GSD region on the sphere with several candidate states in Table \ref{Tab1}. The overlap with the $(112)$, $\overline{1/3}$, and $Z_4$ RR state is always negligible. The overlap with the $(330)$ state is only 0.12 for 12 electrons. The overlap with the I.~L. Pfaffian state is 0.80 for 8 electrons, which is not quite low, but it decays very quickly to 0.73 and 0.66 when the system increases to 10 and 12 electrons.

To gain analytical insight about the numerical results, we also studied the thin torus limit of the coupled $(330)$ system \cite{SM}. We show that at least four of the topological phases conjectured in Table \ref{Tab1} are plausible. Further, we obtain the thin torus patterns of the ground states~\cite{bergholtz2006,seidel2006,ardonne2009,greiter1993}. We then use the resulting patterns to extract several key pieces of information about the nature of the underlying topological order, e.g., GSD, fusion rules, and quantum dimensions of the anyons~\cite{ardonne2009,vaezi2014b}.
In particular we confirmed that the numerically obtained ground states of the 6-fold GSD phase in the thin torus limit are consistent with the thin torus patterns of the bilayer Fibonacci state \cite{SM}.

{\em Discussion.} In summary, we reinvestigated the phase diagram of the coupled $(330)$ bilayer state and considered the effects of the modified interlayer Coulomb interaction by varying its $\mathcal{V}^{\textrm{inter}}_{0}$ and $\mathcal{V}^{\textrm{inter}}_{1}$ components. We have verified the result of McDonald and Haldane~\cite{McDonald1996} indicating that interlayer tunneling and bare Coulomb interaction projected to the LLL can only yield Abelian phases. However, we found that when the pseudopotential components decay smoothly and both $g_1\equiv \mathcal{V}^{\textrm{inter}}_{1}/\mathcal{V}^{\textrm{intra}}_{1}$ and $g_2\equiv\mathcal{V}^{\textrm{inter}}_{1}/\mathcal{V}^{\textrm{inter}}_{0}$ increase, the resulting ground-state exhibits many properties of the bilayer Fibonacci phase. Thus we have provided numerical evidence for the topological quantum phase transition between the $(330)$ state and a NA phase. Moreover the robustness of the bilayer Fibonacci phase which contains Fibonacci anyons which are capable of performing universal quantum computation is tantalizing.

Our result raises the question of how the 6-fold GSD region of our model can be realized experimentally. Our calculations indicate that when $g_1$ and $g_2$ ratios are large enough, the Fibonacci state becomes energetically favorable. If the interlayer distance is small, $\mathcal{V}^{\textrm{inter}}_{1}\approx\mathcal{V}^{\textrm{intra}}_{1}$ so $g_1\approx1$. One strategy to get larger $g_2$ is to go to higher Landau levels with $2/3$ filling~\cite{Kumar2010,Kumar2011}. It is known that in higher Landau levels, the $\mathcal{V}_0$ component of the bare Coulomb interaction decreases with respect to LLL while other components do not change too much. Although in the second Landau level (2LL) for two dimensional electron gas (2DEG) systems, $g_2\approx0.68$ is significantly larger than its value $0.5$ in the LLL, unfortunately we do not have smooth pseudopotentials (the $\mathcal{V}_2$ component is larger than the $\mathcal{V}_1$ component). However, in the 2LL for graphene-like systems, $g_2\approx0.70$ grows and pseudopotentials decay smoothly. In fact, it can be shown through ED that even the bare Coulomb interlayer interaction projected to the 2LL in graphene yields a 6-fold degenerate ground-state, and considering slight finite interlayer distance stabilizes that phase even further~\cite{Future}. Therefore, we expect that graphene is a promising platform for the Fibonacci state.

{\em Note added.} Recently we became aware of related works in Refs.~\cite{Geraedts2015,Peterson2015,Zhu2015}.

{\em Acknowledements.} The authors acknowledge very useful discussions with Nicolas Regnault and Stefanos Kourtis. Z.~L. was supported by the Department of Energy, Office of Basic Energy Sciences, through Grant No.~DE-SC0002140. A.~V., K.~L., and E.-A.~K. were supported by NSF CAREER DMR-0955822.


\begin{thebibliography}{63}
\bibitem{PS}These parameters are chosen to make the ground states deeply in the 6-fold degeneracy region.
$U_0^{\textrm{inter}}=-0.4,U_1^{\textrm{inter}}\gtrsim 0.3$ is already enough to get qualitatively the same results.
\end{thebibliography}
%

\begin{widetext}
\appendix
\newpage
\section*{SUPPLEMENTAL MATERIALS}

In this supplemental material, we study the thin torus limit of our model Hamiltonian for $2/3$ bilayer fractional quantum Hall (FQH) system analytically in section A. We show that when hollow-core interaction dominates, we expect phase transition in the bilayer Fibonacci state. In section B, we present a method based on Ref.~\cite{Li2008} to derive the orbital-cut entanglement spectrum (OES) counting for the bilayer Fibonacci state with an even number of electrons in subregion $A$. A similar calculation would give the counting for other situations and remaining phases. In section C, we compare the particle-cut entanglement spectrum (PES) counting of the ground state in the 6-fold degeneracy region with various candidate states at $\nu=2/3$. Finally in section D, we show the squared wave-function overlap of the ground state in the 6-fold degeneracy region with various candidate states.

\section{A. Thin torus limit of Hamiltonian}

In this section we study the (330) FQH state in the thin torus limit ($L_y/L_x\ll 1$)~\cite{bergholtz2006,seidel2006,ardonne2009} and focus on four regimes, in each of which one interlayer coupling dominates over others and we can ignore other interactions. In the Landau gauge two copies of the Laughlin state at $\nu=1/3$ filling [called as the $(330)$ state] is the exact eigenstate of the following Hamiltonian (usually referred to as $\mathcal{V}_1$ Haldane's pseudo-potential~\cite{Haldane1983}) in the second quantization representation

\begin{eqnarray}
\mathcal{V}_1^{\textrm{intra}}=\sum_{\sigma=\uparrow,\downarrow}\sum_{n,r,s}V^{\para{1}}_{r,s} c^\dag_{n+r,\sigma}  c^\dag_{n+s,\sigma} c_{n,\sigma} c_{n+r+s,\sigma},
\end{eqnarray} where $\sigma=\uparrow,\downarrow$ denotes the layer index, $c_{n,\sigma}$ annihilates an electron with $k_y=2\pi n/L_y$ momentum, $V^{\para{1}}_{r,s}=\frac{\kappa^3}{\sqrt{2\pi}} \para{r^2-s^2}e^{-\kappa^2\frac{r^2+s^2}{2}}$, and $\kappa=2\pi/L_y$.  In this supplementary material, we consider three types of interlayer couplings: (a) {\em Interlayer tunneling} with
\begin{eqnarray}
\mathcal{H}_{t}=-t_{\perp}\sum_{n} c_{n}^\dag \sigma_x c_{n},\quad c_{n}^{\rm T}\equiv\para{c_{n,\uparrow},c_{n,\downarrow}},
\end{eqnarray} coupling Hamiltonian. (b) {\em Interlayer Haldane's $\mathcal{V}_0$ pseudo-potential}:
\begin{eqnarray}
\mathcal{V}_0^{\textrm{inter}}&&=U_0 \sum_{\sigma;n,r,s} V^{\para{0}}_{r,s} c^\dag_{n+r,\sigma} c^\dag_{n+s,-\sigma} c_{n,-\sigma}c_{n+r+s,\sigma} ,\cr
 &&=U_0 \sum_{n,r,s} V^{\para{0}}_{r,s} \hat{s}^\dag_{n+s,n+r} \hat{s}_{n,n+r+s}
\end{eqnarray} where $V^{\para{0}}_{r,s}=\frac{\kappa}{\sqrt{2\pi}} e^{-\kappa^2\frac{r^2+s^2}{2}}$, and $\hat{s}_{n,m}=\frac{1}{\sqrt{2}}\para{c_{n,\uparrow}c_{m,\downarrow}-c_{n,\downarrow}c_{m,\uparrow}}$ is the destruction operator for the (layer-)singlet state between sites $n$ and $m$~(Note that $n$ and $m$ indices are related to the momentum of electrons along $y$ direction. In the Landau gauge, they also dictate the average position of electrons along $x$ axis). In this letter, we consider both repulsive ($U_0>0$) and attractive ($U_0<0$) cases. (c) {\em Hollow-core interaction}, i.e., interlayer Haldane's $\mathcal{V}_1$ pseudo-potential:
\begin{eqnarray}
\mathcal{V}_1^{\textrm{inter}}&&=U_1 \sum_{\sigma; n,r,s} V^{\para{1}}_{r,s} c^\dag_{n+r,\sigma}  c^\dag_{n+s,-\sigma} c_{n,-\sigma} c_{n+r+s,\sigma}\cr
 &&=U_1 \sum_{n,r,s} V^{\para{1}}_{r,s} \hat{t}^\dag_{n+s,n+r} \hat{t}_{n,n+r+s},
 \end{eqnarray} where $\hat{t}_{n,m}=\frac{1}{\sqrt{2}}\para{c_{n,\uparrow}c_{m,\downarrow}+c_{n,\downarrow}c_{m,\uparrow}}$ is the destruction operator for a (layer-)triplet state between sites $n$ and $m$. In the thin torus limit, the parent Hamiltonian as well as interlayer interaction greatly simplify and evaluate to :

\begin{table}[htbp]
  \centering
  \begin{tabular}{@{} |c|c|c| @{}}
  \toprule
  Phase &GSD&Thin torus patterns  \\
    \hline
    $(330)$&9&$[\uparrow\downarrow,0,0],[\uparrow,\downarrow,0],[\downarrow,\uparrow,0]$+translations\\
    $\overline{1/3}$ &3& [$\rightarrow,\rightarrow$,0] + translations\\
    $(112)$&3&$[\frac{\uparrow,\downarrow+\downarrow,\uparrow}{\sqrt{2}},0]$+translations \\
    Bilayer Fibonacci &6& $[\uparrow\downarrow,0,0], [\frac{\uparrow,\downarrow-\downarrow,\uparrow}{\sqrt{2}},0]$+translations   \\
    Interlayer Pfaffian &9& $[\uparrow\downarrow,0,0], ~[\frac{\uparrow,\downarrow-\downarrow,\uparrow}{\sqrt{2}},0],~[\frac{\uparrow,0,\downarrow-\downarrow,0,\uparrow}{\sqrt{2}}]$+translations   \\
    $Z_4$ Read-Rezayi &15&[$\rightarrow,\rightarrow,\rightarrow,\rightarrow,0,0],[\rightarrow,\rightarrow,\rightarrow,0,\rightarrow,0]$,$[\rightarrow,\rightarrow,0,\rightarrow,\rightarrow,0]$+translations\\
    3-Pfaffian state of charge $2e$ bosons &9& $[\uparrow\downarrow,0,0,\uparrow\downarrow,0,0],[\uparrow\downarrow,0,\uparrow\downarrow,0,0,0]$+translations   \\
    \hline
  \end{tabular}
  \caption{Ground-state degeneracy and the thin torus patterns associated with some of the candidate phases at $\nu=2/3$. The number of different patterns for each phase equals the corresponding GSD. In the thin torus limit, ground-state is represented by repetition of either patterns. The resulting patterns encode the spatial density of electrons. Electrons with $\uparrow (\downarrow)$ label live on top (bottom) layer. $\rightarrow \equiv \frac{1}{\sqrt{2}}(\uparrow+\downarrow)$ represents resonating electrons.}
  \label{Tab2}
\end{table}

\begin{eqnarray}
 \mathcal{V}_1^{\textrm{intra}}\to \sum_{n}\sum_{r=1,2}V^{\para{1}}_{r,0} \hat{n}_{n+r,\sigma} \hat{n}_{n,\sigma},\label{eq:V1intra}
\end{eqnarray}
\begin{eqnarray}
 \mathcal{V}_0^{\textrm{inter}}\to && U_0 \sum_{n} \left[V^{\para{0}}_{0,0} \hat{n}_{n,\uparrow} \hat{n}_{n,\downarrow}+V^{\para{0}}_{1,0} \hat{s}^\dag_{n,n+r} \hat{s}_{n,n+r}\right],\label{eq:V0inter}
 \end{eqnarray}
\begin{eqnarray}
 \mathcal{V}_1^{\textrm{inter}}\to U_1 \sum_{n} V^{\para{1}}_{1,0} \hat{t}^\dag_{n,n+1} \hat{t}_{n,n+1},\label{eq:V1inter}
 \end{eqnarray}
where $\hat{n}_{n,\sigma}=c_{n,\sigma}^\dag c_{n,\sigma}$ is the electron number operator at site $n$.
$\mathcal{V}_1^{\textrm{intra}}$ is already diagonal in the occupation basis and can be easily solved. $\mathcal{V}_1^{\textrm{intra}}$ for spin up sector contains three ground-states: $\ket{\uparrow,0,0,\uparrow,0,0,\uparrow,0,0,\cdots}$, $\ket{0,\uparrow,0,0,\uparrow,0,0,\uparrow,0,\cdots}$, $\ket{0,0,\uparrow,0,0,\uparrow,0,0,\uparrow,\cdots}$. Similarly the spin down component of $\mathcal{V}_1^{\textrm{intra}}$ can be solved and the ground-state of full intra-layer Hamiltonian is simply the 9 different possible tensor products of the ground-states of the parts. (see Table \ref{Tab2}).

Below we study the four extreme limits:

\noindent (i) $\abs{t_{\perp}} \gg \abs{U_0}, U_{1}$: In this limit, we must first diagonalize $\mathcal{H}_{t}$ , and then consider the effect of intra-layer repulsion. In order to minimize $\mathcal{H}_{t}$, we need to project spin along $+x$ direction (if $t$ is a real positive number), i.e. electrons must occupy the symmetric state $\ket{\rightarrow}_i=\frac{1}{\sqrt{2}}\para{\ket{\uparrow}_i+\ket{\downarrow}_i}$. Then there are two regimes:

(a) $t_{\perp}$ is much larger than $V^{(1)}_{1,0}$, then we obtain the following ground-state patterns:
\begin{eqnarray}
&&[\rightarrow,\rightarrow,0] \equiv \ket{\rightarrow,\rightarrow,0,\rightarrow,\rightarrow,0,\rightarrow,\rightarrow,0,\cdots},\cr
&&[0,\rightarrow,\rightarrow]\equiv \ket{0,\rightarrow,\rightarrow,0,\rightarrow,\rightarrow,0,\rightarrow,\rightarrow,\cdots},\cr
&&[\rightarrow,0,\rightarrow]\equiv \ket{\rightarrow,0,\rightarrow,\rightarrow,0,\rightarrow,\rightarrow,0,\rightarrow,\cdots}.
\end{eqnarray}
By consulting Table \ref{Tab2}, we conclude that for very extreme interlayer tunneling we expect the particle-hole conjugate of the 1/3 Laughlin state $\overline{1/3}$ state.

(b) $t_{\perp}$ is comparable or less than $V^{(1)}_{1,0}$, then using second order perturbation we obtain the following ground-state patterns~\cite{vaezi2014b}:
\begin{eqnarray}
&&[\frac{1}{\sqrt{2}}\para{\uparrow,\downarrow+\downarrow,\uparrow},0] +\mbox{two other translations},
\end{eqnarray}
where as before [a,b,c] denotes the building block (unit cell) of the ground-state. The ground-states can be obtained by repeating  these unit cells all over 1D lattice associated with occupation number basis. Table \ref{Tab2} suggests it is $(112)$ Halperin state~\cite{Halperin1984}.

\noindent (ii) $U_{1} \gg \abs{t_{\perp}}, \abs{U_0}$: In this case, we again must deal with $\mathcal{V}_1^{\textrm{inter}}$. Since, it consists of triplet counting operators, the ground-state must be spin singlet between near neighbors. Doubly occupied sites are also spin singlet. Therefore, the thin torus patterns of the ground-state patterns are:

\begin{eqnarray}
&&[200]\equiv [\uparrow\downarrow,0,0],\quad [110]\equiv [\frac{1}{\sqrt{2}}\para{\uparrow,\downarrow-\downarrow,\uparrow},0] +\mbox{translations}.\cr
&&
\end{eqnarray}
In this way, we obtain six degenerate ground-states. The thin torus patterns tells us that every three consecutive orbitals contain exactly three orbital in the ground-state. According to table \ref{Tab2} it represents the bilayer Fibonacci state which contains Fibonacci NA anyons.

\noindent (iii) $U_0\gg \abs{t_{\perp}}, U_1$: Considering the onsite term in $\mathcal{V}_0^{\textrm{inter}}$, because of strong repulsion between two layers, ground-state cannot contain doubly occupied sites. This leaves us with six low energy patterns. However, taking the nearest neighbor interaction, i.e. $U_0V_{1,0}\hat{s}^\dag _{i,i+1}\hat{s}_{i+1}$ favors triplet formation. Therefore, we expect the following three ground-states:

\begin{eqnarray}
&&[\frac{1}{\sqrt{2}}\para{\uparrow,\downarrow+\downarrow,\uparrow},0] +\mbox{translations}.
\end{eqnarray}
According to table \ref{Tab2} the above thin torus patterns are signatures of (112) Abelian state.

\noindent (iv) $-U_0\gg \abs{t_{\perp}}, U_1$: When interlayer coupling is attractive and strong, we expect electrons on the same form a bound-state. These bound-states can be imagined as charge $-2e$ bosons. Assuming these electrons pairs are localized and tightly bound, we can project the remaining terms in the Hamiltonian to the this low energy subspace and obtain an effective interaction, between bosons. Doing so, we obtain the following effective Hamiltonian:

 \begin{eqnarray}
&& H_{\rm eff}\to \sum_{n}\sum_{r=1,2}\para{V^{\para{1}}_{r,0}-\abs{U^0}V^{\para{0}}_{r,0}}\hat{n}^b _{n+r} \hat{n}^{b}_{n},\label{eq:Heff}
\end{eqnarray}
where $\hat{n}^b_i=b_i^\dag b_i$ is the density of bosons (electron pairs). Now consider the following two groups of patterns:

\begin{eqnarray}
(a):~ &&[\uparrow\downarrow,0,0,\uparrow\downarrow,0,0]+\mbox{translations},\cr
(b):~&& [\uparrow\downarrow,0,\uparrow\downarrow,0,0,0]+\mbox{translations}.
\end{eqnarray}
These two groups are not exactly degenerate with respect to the effective Hamiltonian Eq. \eqref{eq:Heff} in the thin torus limit. However, by considering appropriate three-body interactions to the effective Hamiltonian Eq. \eqref{eq:Heff}, the two groups above can become degenerate and evolve together. In that case, the resulting Hamiltonian describes nine degenerate ground-state patterns. These patterns are indeed root configurations of the ground-state wave function. Comparing the above nine patterns with those in Table \ref{Tab2}, we conclude that the above phase with additional three-body interactions is described by the 3-Pfaffian state.

\section{B. Edge state counting using the thin torus patterns}

In this section we closely follow the method developed in Ref. \cite{Li2008} to compute the edge state or entanglement counting of the bilayer Fibonacci state. The counting of other states can be found similarly. For this end, let us consider the half infinite cylinder geometry with a boundary at $x=0$. In the thin cylinder limit, one of the ground-states of the bilayer Fibonacci state is the following pattern (this is the unique ground-state pattern on sphere geometry):

\begin{eqnarray}
&&\ket{\rm GS}_1=\cdots \uparrow\downarrow,0,0,\uparrow\downarrow,0,0,\uparrow\downarrow,0,0,\uparrow\downarrow,0,0,0,0,0,0,0\cdots\label{eq:GS}
\end{eqnarray}
On the left side of $x=0$, the $\para{\uparrow\downarrow,0,0}$ pattern is repeated all the way to $-\infty$, while there is no electron on its right side. The total momentum (or total angular momentum on sphere geometry) of an arbitrary state is defined as:
\begin{eqnarray}
&&L=\sum_{i\in Z}\hat{n}_{i}i, \quad \hat{n}_i=\sum_{\sigma}c_{i,\sigma}^\dag c_{i,\sigma}.
\end{eqnarray}
Let us define $l$ as the relative momentum of an arbitrary state with respect to the ground-state in Eq. \eqref{eq:GS}, i.e. $l=L-L_{\rm GS}$. For example, $l=0$ for the ground-state itself. Now we want to study excited states with relative angular momentum $l$. There are two categories of excitations:

(a) Edge excitations: These are low energy excitations which respect the generalized Pauli exclusion rules for the bilayer Fibonacci state.

(b) Bulk excitations: Any state which does not respect the generalized Pauli exclusion rules.

The Pauli exclusion rule for the bilayer Fibonacci state is:  (i) In every three consecutive orbitals, there should be at most two electrons. (ii) Electrons with distance one or two (e.g. 110 or 101 configurations) from singlet pairs and must be counted once. The formation of 110 singlets has priority unless the formation of 101 singlets can increase the total number of singlets.  (iii) Electrons with distance more than two (e.g. 1001 or 10001) can form both singlet and triplet states and thus must be counted twice.

In the following, we use the generalized Pauli exclusion rule mentioned above to find the counting of the edge excitations for $l=1,2,3,4$.

{\bf Example 1. $l=1$:} There is only one admissible edge excitation (the states in this section need to be normalized):
\begin{eqnarray}
&&\ket{l=1,{\rm edge}}_1=\cdots \uparrow\downarrow,0,0,\uparrow\downarrow,0,0,\uparrow\downarrow,0,0,\para{\uparrow,\downarrow-\downarrow,\uparrow}{},0,0,0,0\cdots
\end{eqnarray}

{\bf Example 2. $l=2$:} There are three admissible edge excitations:
\begin{eqnarray}
&&\ket{l=2,{\rm edge}}_1=\cdots\uparrow\downarrow,0,0,\uparrow\downarrow,0,0,\uparrow\downarrow,0,0,0,\uparrow\downarrow,0,0,0,0,\cdots \cr
&&\ket{l=2,{\rm edge}}_2=\cdots\uparrow\downarrow,0,0,\uparrow\downarrow,0,0,0,\para{\uparrow,\downarrow-\downarrow,\uparrow}{},0,\para{\uparrow,\downarrow-\downarrow,\uparrow}{},0,0,0,0,\cdots\cr
&&\ket{l=2,{\rm edge}}_3=\cdots\uparrow\downarrow,0,0,\uparrow\downarrow,0,0,\uparrow\downarrow,0,0,\para{\uparrow,0,\downarrow-\downarrow,0,\uparrow}{},0,0,0\cdots,
\end{eqnarray}

{\bf Example 3. $l=3$: }There are five admissible edge excitations:
\begin{eqnarray}
&&\ket{l=3,{\rm edge}}_1=\cdots\uparrow\downarrow,0,0,\uparrow\downarrow,0,0,\uparrow\downarrow,0,0,0,\para{\uparrow,\downarrow-\downarrow,\uparrow}{},0,0,0,\cdots \cr
&&\ket{l=3,{\rm edge}}_2=\cdots\uparrow\downarrow,0,0,\uparrow\downarrow,0,0,\para{\uparrow,\downarrow-\downarrow,\uparrow}{},0,0,\uparrow\downarrow,0,0,0,0,\cdots\cr
&&\ket{l=3,{\rm edge}}_3=\cdots\uparrow\downarrow,0,0,\uparrow\downarrow,0,0,\para{\uparrow,\downarrow-\downarrow,\uparrow}{},0,\para{\uparrow,0,\downarrow-\downarrow,0,\uparrow}{},0,0,0,\cdots \cr
&&\ket{l=3,{\rm edge}}_4=\cdots\uparrow\downarrow,0,0,\para{\uparrow,\downarrow-\downarrow,\uparrow}{},0,\para{\uparrow,\downarrow-\downarrow,\uparrow}{},0,\para{\uparrow,\downarrow-\downarrow,\uparrow}{},0,0,0,0,\cdots \cr
&&\ket{l=3,{\rm edge}}_5=\cdots\uparrow\downarrow,0,0,\uparrow\downarrow,0,0,\uparrow\downarrow,0,0,\para{\uparrow,0,0,\downarrow-\downarrow,0,0,\uparrow}{},0,0,\cdots.\\
&&\ket{l=3,{\rm bulk}}_6=\cdots\uparrow\downarrow,0,0,\uparrow\downarrow,0,0,\uparrow\downarrow,0,0,\para{\uparrow,0,0,\downarrow+\downarrow,0,0,\uparrow}{},0,0,\cdots.
\end{eqnarray}

\begin{figure}
\centerline{\includegraphics[width=0.6\linewidth] {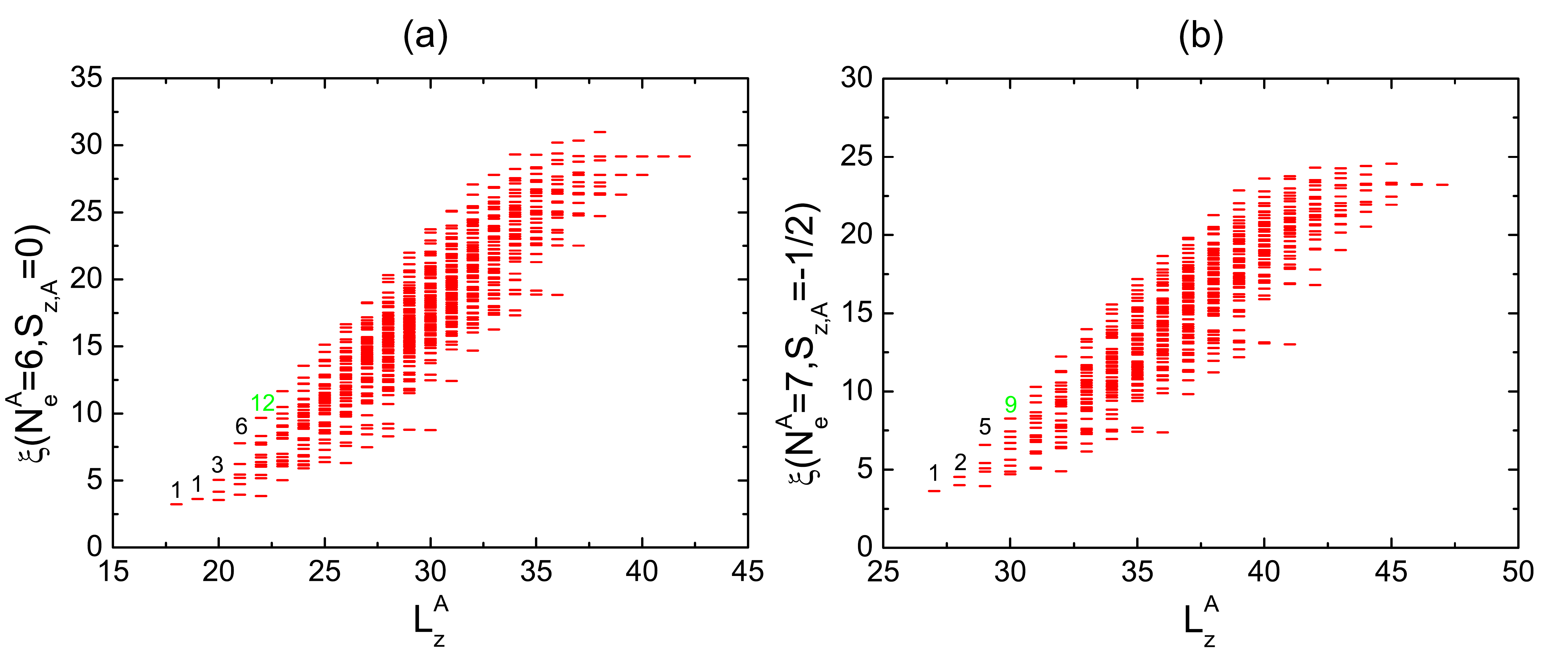}}
\caption{(Color online) The OES of $N_e=14$ electrons in the $N_e^A=6,S_z^A=0$ and $N_e^A=7,S_z^A=-1/2$ sectors in the limit of $V_1^{\textrm{inter}}\gg V_1^{\textrm{intra}}$ and $t_{\perp}=0$. The countings that suffer from finite-size effect are labeled by green numbers.}
\label{fg:oesdas}
\end{figure}

{\bf Example 4. $l=4$: }There are ten admissible edge excitations:\begin{eqnarray}
&&\ket{l=4,{\rm edge}}_{1~}=\cdots\uparrow\downarrow,0,0,\uparrow\downarrow,0,0,\uparrow\downarrow,0,0,0,0,{\uparrow\downarrow},0,0,0,\cdots \cr
&&\ket{l=4,{\rm edge}}_{2~}=\cdots\uparrow\downarrow,0,0,\uparrow\downarrow,0,0,0,\uparrow\downarrow,0,0,\uparrow\downarrow,0,0,0,0,\cdots\cr
&&\ket{l=4,{\rm edge}}_{3~}=\cdots\uparrow\downarrow,0,0,\uparrow\downarrow,0,0,\uparrow\downarrow,0,0,0,\para{\uparrow,0,\downarrow-\downarrow,0,\uparrow}{},0,0,\cdots \cr
&&\ket{l=4,{\rm edge}}_{4~}=\cdots\uparrow\downarrow,0,0,\para{\uparrow,\downarrow-\downarrow,\uparrow}{},0,\para{\uparrow,\downarrow-\downarrow,\uparrow}{},0,0,\uparrow\downarrow,0,0,0,0,\cdots \cr
&&\ket{l=4,{\rm edge}}_{5~}=\cdots\uparrow\downarrow,0,0,\para{\uparrow,\downarrow-\downarrow,\uparrow},0,\para{\uparrow,\downarrow-\downarrow,\uparrow},0,\para{\uparrow,0,\downarrow-\downarrow,0,\uparrow},0,0,0,\cdots\cr
&&\ket{l=4,{\rm edge}}_{6~}=\cdots\uparrow\downarrow,0,0,\uparrow\downarrow,0,0,\para{\uparrow,\downarrow-\downarrow\uparrow},0,0,\para{\uparrow,\downarrow-\downarrow\uparrow},\cdots\cr
&&\ket{l=4,{\rm edge}}_{7~}=\cdots\uparrow\downarrow,0,0,\uparrow\downarrow,0,0,\uparrow,\uparrow,0,0,\downarrow,\downarrow,\cdots\cr
&&\ket{l=4,{\rm edge}}_{8~}=\cdots\uparrow\downarrow,0,0,\uparrow\downarrow,0,0,\para{\uparrow,0,\downarrow-\downarrow,0,\uparrow},\para{\uparrow,0,\downarrow-\downarrow,0,\uparrow},0,0,0,\cdots\cr
&&\ket{l=4,{\rm edge}}_{9~}=\cdots\uparrow\downarrow,0,0,\uparrow\downarrow,0,0,\para{\uparrow,\downarrow-\downarrow,\uparrow},0,\para{\uparrow,0,0,\downarrow-\downarrow,0,0,\uparrow},0,0,\cdots\cr
&&\ket{l=4,{\rm bulk}}_{10~}=\cdots\uparrow\downarrow,0,0,\uparrow\downarrow,0,0,\para{\uparrow,\downarrow-\downarrow,\uparrow},0,\para{\uparrow,0,0,\downarrow+\downarrow,0,0,\uparrow},0,0,\cdots\cr
&&\ket{l=4,{\rm edge}}_{11}=\cdots\uparrow\downarrow,0,0,\uparrow\downarrow,0,0,\uparrow\downarrow,0,0,\para{\uparrow,0,0,0,\downarrow-\downarrow,0,0,0,\uparrow},0,\cdots \cr
&&\ket{l=4,{\rm bulk}}_{12}=\cdots\uparrow\downarrow,0,0,\uparrow\downarrow,0,0,\uparrow\downarrow,0,0,\para{\uparrow,0,0,0,\downarrow+\downarrow,0,0,0,\uparrow},0,\cdots\cr
&&\ket{l=4,{\rm edge}}_{13~}=\cdots\para{\uparrow,\downarrow-\downarrow,\uparrow},0,\para{\uparrow,\downarrow-\downarrow,\uparrow},0,\para{\uparrow,\downarrow-\downarrow,\uparrow},0,\para{\uparrow,\downarrow-\downarrow,\uparrow},0,0,0,0,\cdots
\end{eqnarray}
Note that $\ket{l=4,{\rm edge}}_{13~}$ can happen only for $N_e^A\geq 8$, and thus the counting for $N_e^A=6$ and $l=4$ is 12 rather than 13. Similarly, we can find the counting for higher angular momenta in the $N_e^A=0$ (mod 2) and $S_z^A=0$ sector. The OES counting for other values of $N_e^A$ modulo 2 and $S_z^A$ can also be obtained with possible slight changes of the exclusion rule. In the limit of $V_1^{\textrm{inter}}\gg V_1^{\textrm{intra}}$ and $t_{\perp}=0$, the OES counting is very clean (Fig.~\ref{fg:oesdas}).

\begin{table}[htbp]
  \centering
  \begin{tabular}{@{} |c|c|c|c| @{}}
  \toprule
  Phase &$N_e=8,N_e^A=4$&$N_e=10,N_e^A=4$&$N_e=12,N_e^A=6$  \\
    \hline
    $(330)$&784&3025&48400\\
    $(112)$&784&3025&48400\\
    Interlayer Pfaffian &1125&6085&104491   \\
    $Z_4$ Read-Rezayi&210&715&7506\\
    6-fold degenerate state&505&2242&32840   \\
    \hline
  \end{tabular}
  \caption{The PES counting of the ground state in the 6-fold degeneracy region as well as various candidate states on the sphere for different system sizes in the $S_z^A=0$ sector.}
  \label{Tab3}
\end{table}

\begin{table}[htbp]
  \centering
  \begin{tabular}{@{} |c|c|c|c|c| @{}}
  \toprule
  Phase &$N_e=6$&$N_e=8$&$N_e=10$&$N_e=12$  \\
    \hline
    $(330)$&0.44&0.28&0.18&0.12\\
    Interlayer Pfaffian &0.89&0.80&0.73&0.66   \\
    \hline
  \end{tabular}
  \caption{The squared wave function overlap of the ground state in the 6-fold degeneracy region with various candidate states on the sphere. We use the Hamiltonian $H_{\textrm{coulomb}}^{\textrm{intra}}+H_{\textrm{coulomb}}^{\textrm{inter}}+U_0^{\textrm{inter}}\mathcal{V}_0^{\textrm{inter}}+U_1^{\textrm{inter}}\mathcal{V}_1^{\textrm{inter}}$ with $U_0^{\textrm{inter}}=-0.4$ and $U_1^{\textrm{inter}}=0.6$ to get the ground state in the 6-fold degeneracy region. We use $H=\mathcal{V}_1^{\textrm{intra}}+\mathcal{V}_1^{\textrm{inter}}+\mathcal{H}_{t}$ with $t_{\perp}=2$ to get the $\overline{1/3}$ state. We antisymmetrize the (330) state to get the $Z_4$ state. Other states are generated from their parent Hamiltonians. The overlap with the $(112)$, $\overline{1/3}$ and $Z_4$ RR state is always negligible.}
  \label{Tab4}
\end{table}

\section{C. Quasihole counting using the PES}
In this section we show the PES counting of the ground state in the 6-fold degeneracy region as well as various candidate states on the sphere (Table \ref{Tab3}). We can see that the counting of the ground state in the 6-fold degeneracy region is very different from all candidate states that we have checked, thus ruling their possibilities out.

\section{D. Wave function overlap}
We have computed the squared wave function overlap of the ground state in the 6-fold degeneracy region with various candidate states on the sphere (Table \ref{Tab4}). The overlap with the $(112)$, $\overline{1/3}$ and $Z_4$ RR state is always negligible. From the aspect of the overlap, we can see that the main competing phase is the inter-layer Pfaffian. However, the ground-state degeneracy on the torus as well as OES and PES can rule it out.

\end{widetext}

\end{document}